\documentclass[conference]{IEEEtran}
\IEEEoverridecommandlockouts
\usepackage{cite}
\usepackage{amsmath,amssymb,amsfonts}
\usepackage{algorithmic}
\usepackage{graphicx}
\usepackage{textcomp}
\usepackage[dvipsnames]{xcolor}
\usepackage[utf8]{inputenc}
\usepackage{makecell}

\def\cbrace{\iffalse{\else}\fi}

\begin{document}

\title{In-Memory Resistive RAM Implementation \\ of Binarized Neural Networks \\ for Medical Applications 
\thanks{\textbf{This work was supported by the European Research Council Grant NANOINFER (715872) and  ANR grant NEURONIC (ANR-18-CE24-0009).}}
}

\author{\IEEEauthorblockN{
Bogdan Penkovsky\IEEEauthorrefmark{1},  
Marc Bocquet\IEEEauthorrefmark{2}, 
Tifenn Hirtzlin\IEEEauthorrefmark{1},
Jacques-Olivier Klein\IEEEauthorrefmark{1}, \\
Etienne Nowak\IEEEauthorrefmark{3}, 
Elisa Vianello\IEEEauthorrefmark{3},
Jean-Michel Portal\IEEEauthorrefmark{2} 
and 
Damien Querlioz\IEEEauthorrefmark{1}}
\IEEEauthorblockA{\IEEEauthorrefmark{1}C2N, Universit\'e Paris-Saclay, CNRS, Palaiseau, France.
Email: damien.querlioz@c2n.upsaclay.fr}
\IEEEauthorblockA{\IEEEauthorrefmark{2}IM2NP, Aix-Marseille Université, Université de Toulon, CNRS, Marseille, France
}\IEEEauthorblockA{
\IEEEauthorrefmark{3}CEA, LETI, Grenoble, France.}
}

\maketitle

\begin{abstract}
The advent of deep learning has considerably accelerated machine learning development. The deployment of deep neural networks at the edge is however limited by their high memory and energy consumption requirements. With new memory technology available, emerging Binarized Neural Networks (BNNs) are promising to reduce the energy impact of the forthcoming machine learning hardware generation, enabling machine learning on the edge devices and avoiding data transfer over the network. In this work, after presenting our implementation employing a hybrid CMOS - hafnium oxide resistive memory technology, we suggest strategies to apply BNNs to biomedical signals such as electrocardiography and electroencephalography, keeping accuracy level and reducing memory requirements.  We investigate the memory-accuracy trade-off
when binarizing whole network and binarizing solely the classifier part.
We also discuss how these results translate to the edge-oriented Mobilenet~V1 neural network on the Imagenet task.
The final goal of this research is to enable smart autonomous healthcare devices.
\end{abstract}

\section{Introduction}

With recent advances in machine learning, multiple challenging tasks
are now solved by computers with  human-level accuracy \cite{Esteva2017}, or even better \cite{silver2016mastering}, and smart assistance services  are available for anyone. However, 
those services operate in the {cloud},
requiring energy-expensive data transfer over the network.
We envision that the quality of medical services can be substantially
improved with the use of machine learning, with applications such as
stroke and heart attack prevention, epileptic seizure prediction,
post-hospital monitoring and rehabilitation, and brain-computer interfaces for people with disabilities.
Those  services should
be available at the edge to ensure privacy, security, and low latency.
This mobility rises new challenges:
maximizing battery life and making hardware
as small and lightweight as possible.

The major drain of energy
in modern digital electronics, especially when performing data-intensive artificial intelligence tasks, comes from data shuffling
between processing logic and memory \cite{pedram2017dark}.
This issue can be substantially alleviated by applying
in-memory computing principles, 
which eliminate the von Neumann bottleneck and are especially applicable
for neural network architectures.
This idea is particularly relevant today, with the emergence of novel non-volatile memory technologies that are fully compatible with modern CMOS processes and appear ideal for the in-memory implementation of neural networks
\cite{yu2018neuro,indiveri2015memory}.
An obvious drawback of this approach
is the limited amount of on-chip memory.
Indeed, when talking about in-memory
computing we cannot rely on external memories,
thus adhering to the amount of on-chip
memory becomes critical.
Unfortunately, deep neural networks often require
considerable amounts of memory  \cite{Siu2018}.

Significant efforts have been made to reduce the memory footprint of neural networks.
{Compact network design} aims at reducing the number of neural network
parameters and operations, while maintaining accuracy \cite{Iandola2016, Howard2017, Zhang2018}.
These networks still use real-valued weights and activations represented
as 32-bits floating point numbers.
{Quantized neural networks} aim at reducing memory requirement by reducing
the number of bits  in weights and activations \cite{hubara2017quantized}.
Eight-bit quantization is particularly successful in applications, as it usually requires no retraining \cite{jouppi2017datacenter}.
{Binarized neural networks} (BNNs) are the extreme evolution of quantized networks with a precision reduced to a single bit \cite{courbariaux2016binarized, rastegari2016xnor}. Beyond weight and activation, the memory footprint can also be reduced with  binary representation of the inputs using stochastic sampling \cite{Hirtzlin2019b}.

In this work, we address the energy efficiency
challenge by proposing an in-memory implementation of binarized neural network using emerging memories.
We realize that bit errors inherent to resistive memory technologies are a challenge and introduce a two transistor - two resistor approach to mitigate the issue.
We validate our approach, which is adaptable to different emerging memory technologies, using measurements on a fabricated hybrid CMOS/resistive memory chip \cite{bocquet2018,hirtzlin2019digital}.
We  evaluate
our all-binarized convolutional neural network approach on medical time-signals such as electroencephalogram 
and electrocardiogram. 
We suggest that partial binarization
may sometimes be an option
to minimize the accuracy loss compared to original
neural networks with real weights, while substantially reducing
memory requirements.
Finally, we speculate about the application
of the same partial binarization
strategy for generic computer vision, which could have direct relevance to medical imaging.

\begin{figure}[ht]
    \centering
    \includegraphics[width=0.42\textwidth]{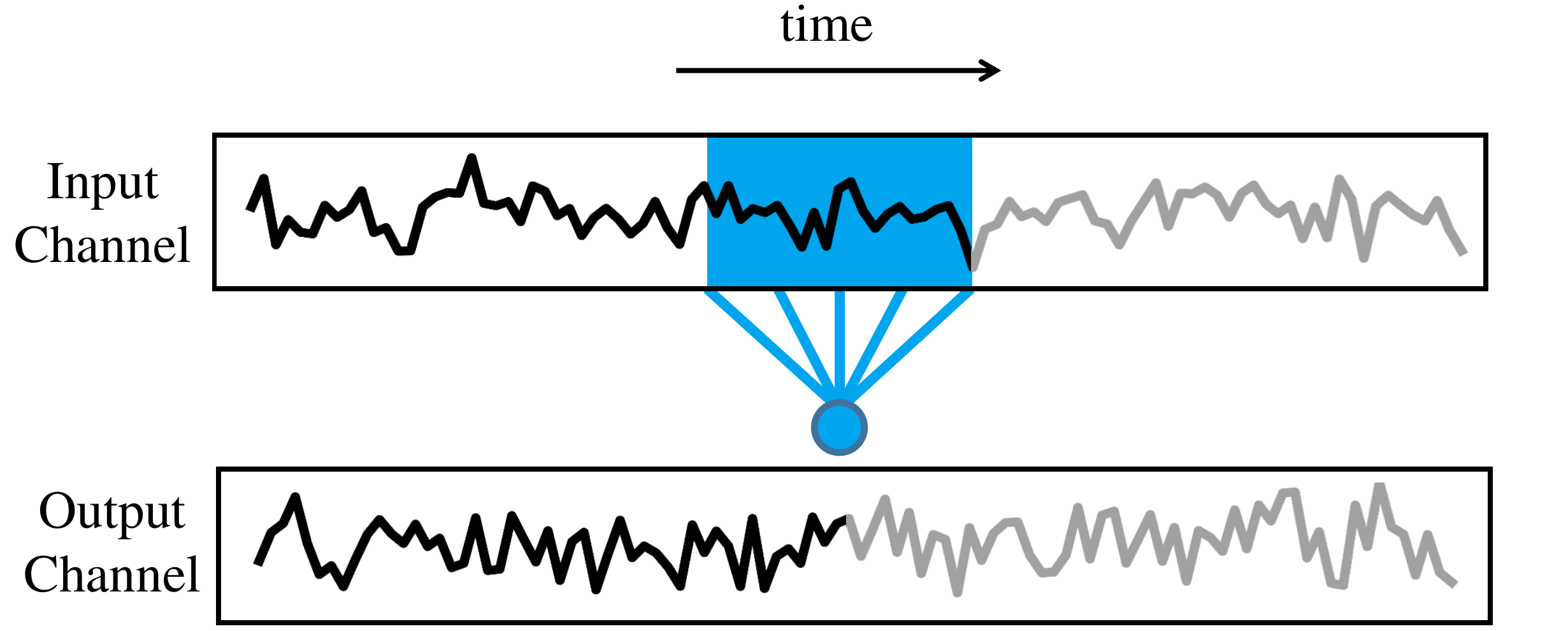}
    \caption{Illustration of a temporal 1D convolution.}
    \label{fig:Temporal convolution}
\end{figure}

\section{In-Memory Implementation of Binarized Neural Networks}

\subsection{Binarization of Neural Networks}

Convolutional neural networks (CNNs)
are a state-of-the-art supervised deep learning
architecture \cite{lecun2015deep}, which 
consists of  a {feature extractor}
and a {classifier}.
The feature extractor has multiple
sparsely-connected convolutional layers,
whereas the classifier exhibits
dense, all-to-all connection topology in its layers.
In both cases the elementary network unit, the artificial neuron, performs a nonlinear transformation $f$
over a dot product between an input vector $\textbf{x}$
and learned weights $\textbf{w}$:
\begin{equation}
y = f\left(\sum_j w_j x_j + b\right),
\label{eq:neuron}
\end{equation}
where  $b$ is a bias term.
Neural networks are trained using a gradient-descent family optimization method to minimize the error
between their actual and desired outputs.
Convolutional layers connection topology is based on
the discrete convolution transformation between
an input tensor $\textbf{x}$ and a sliding window $\textbf{w}$,
called  convolution {kernel}.
A discrete convolution in the 1-D case (Fig.~\ref{fig:Temporal convolution}) is defined as
\begin{equation}
\left( \textbf{x} * \textbf{w} \right)_i = \sum_{m=0}^{D_K-1} w_{i-m} x_m,
\label{eq:conv}
\end{equation}
 where
$\textbf{w} \in \mathbb{R}^{D_{K}}$, $\textbf{x} \in \mathbb{R}^{D_{F}}$, and $0 \le i < D_K + D_F - 1$. 
This operation
can be extended to multiple dimensions in a straightforward manner.
Convolutional neural networks also feature {pooling} layers,
to reduce the spatial resolution of layers, therefore reducing
the number of subsequent computations. 
Finally, the classifier discriminates outcomes
between the known classes, based on features forwarded from the
convolutional layers.

Considerable work has investigated the implementation of convolutional neural networks  in hardware using emerging memory, with architectures such as ISAAC \cite{shafiee2016isaac} or PRIME \cite{chi2016prime}, using either digital or analog coding for  the weights. Digital coding  has substantial memory requirements. Analog coding, by contrast, requires only two devices per weight (two devices are needed for the possibility to store negative weights), but has the disadvantage of requiring complex peripherals such as analog-to-digital and digital-to-analog converters with their associated high area overhead \cite{shafiee2016isaac,chi2016prime}. 

Binarizing the neural network provides an alternative route to reducing the area and energy consumption of hardware neural networks.
In binarized neural networks (BNNs), we use weights with values of either $+1$ or $-1$, 
and the $\textrm{sign}$ function as activation function.
This allows  simplifying Eq.~\eqref{eq:neuron} to
\begin{equation}
    y = \textrm{sign}\left( \textrm{popcount}\left( \textrm{XNOR}(w_j, x_j) \right) - b\right),
\end{equation}
where $\textrm{popcount}$ is a function counting the number of $1$ bits
and $b$ is a learned neuron threshold.
On top of the low memory requirements of BNNs, replacing multiplication circuits with simple $\textrm{XNOR}$ logic gates can greatly reduce the circuit area.

\subsection{Implementing Binarized Neural Networks}

\begin{figure}
    \centering
    \includegraphics[width=3.6in]{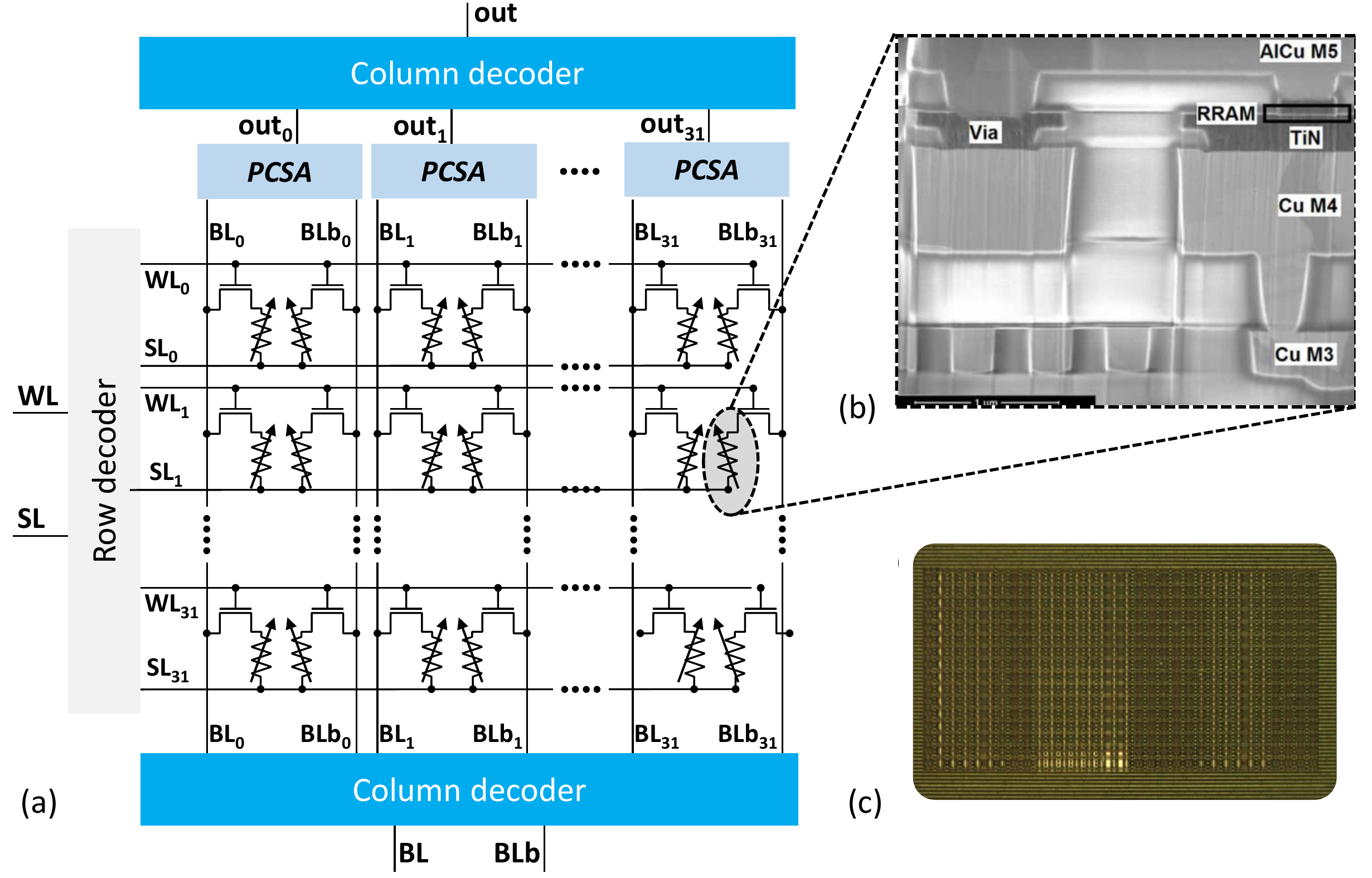}
    \caption{(a) Schematic of a 1K synapses RRAM cells (b) Scanning Electron Microscopy image of an RRAM device integrated in the BEOL of our technology (c) Photograph of the die of our test chip with 1K synapses / 2K RRAM cells.}
    \label{fig:die}
\end{figure}

\begin{figure}
    \centering
    \includegraphics[width=3.4in]{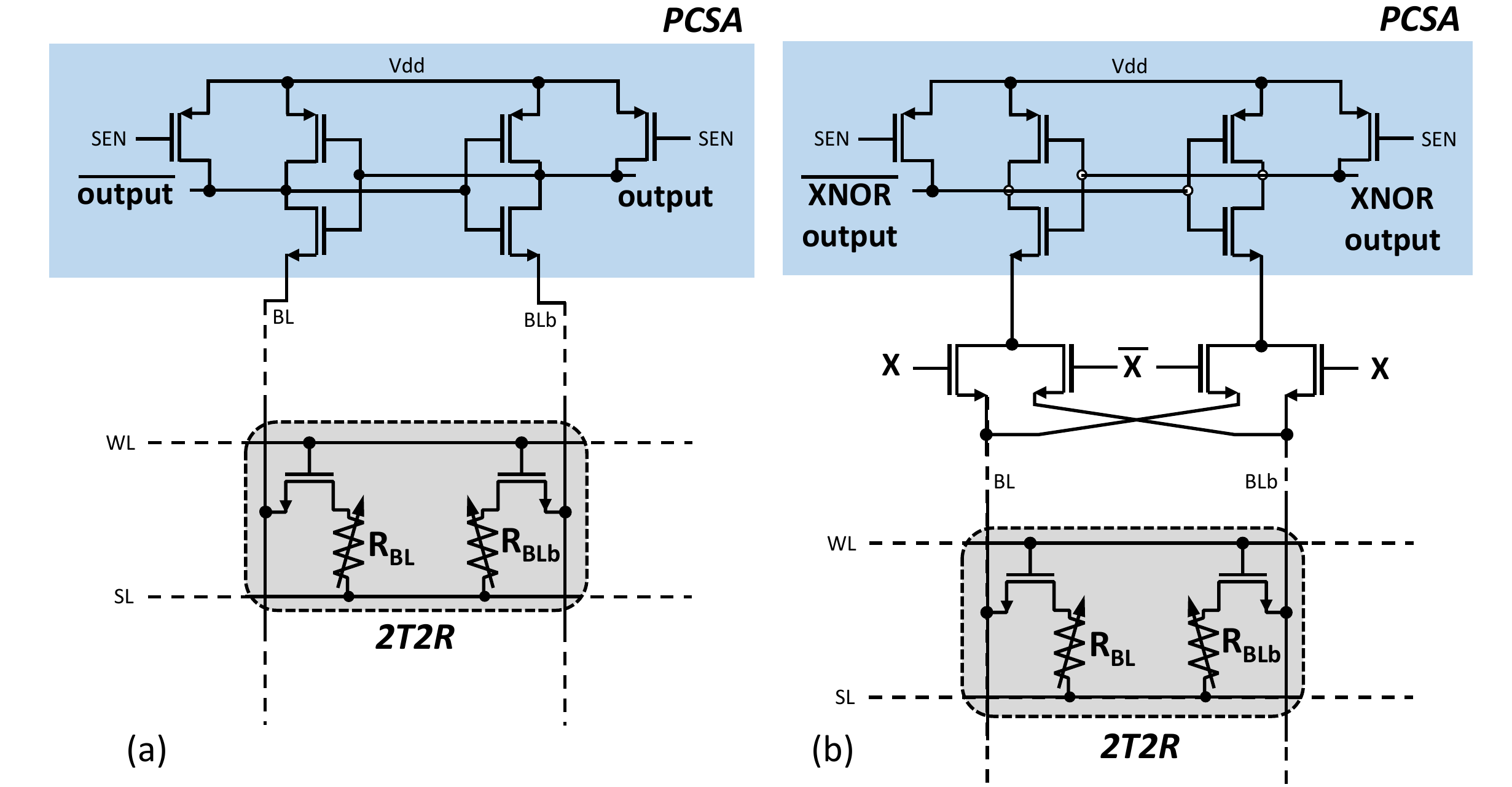}
    \caption{(a) Schematic of the sense amplifier used to extract the binary weight from a 2T2R synapse. (b) Version augmented with an XNOR feature.}
    \label{fig:PCSA}
\end{figure}

We now introduce our technique for the energy-efficient in-memory implementation of binarized neural networks exploiting resistive memory.
Our test chip uses hafnium oxide-based resistive memory, fully integrated within the back end of line (BEOL) of a commercial  130~nanometer CMOS process (Fig.~\ref{fig:die}(b)).
A photograph of our die is presented in Fig.~\ref{fig:die}(c) and its simplified schematic  in Fig.~\ref{fig:die}(a).
The considerable challenge to implement in-memory computing with resistive memory is their inherent device variation, which leads to bit errors \cite{ielmini2018memory,ly2018role}.
Conventional digital designs suppress these errors by relying on multiple error correcting codes (ECC). However, the use of ECC in the context of in-memory binarized neural network is not satisfying:
the computation of error detection and correction is more complicated than the one of binarized neural network, and would dominate area and energy consumption. 
Additionally, ECC goes against the idea of integrating part of the computation within the memory array or sensing circuit, a major idea of in-memory computing.
Our design therefore uses an alternative ECC-less approach to reduce the number of bit errors, by relying on a two transistor/two resistor (2T2R) architecture (Fig.~\ref{fig:die}(a)), where synaptic weights are stored in a differential fashion: by convention, a device pair programmed in the low resistance/high resistance state means a synaptic weight of $+1$, and reciprocally a pair programmed in the high resistance/low resistance state  means a synaptic weight $-1$.
Precharge sense amplifiers (PCSA, Fig.~\ref{fig:PCSA}(a)) are used to compare the resistance states of the two devices of a pair, and therefore read the synaptic weight.
An attractive possibility of the approach is the option to incorporate the XNOR operation of BNNs directly within the precharge sense amplifier, by the addition of solely four transistors  (Fig.~\ref{fig:PCSA}(b)).

\begin{figure}
    \centering
    \includegraphics[width=3.3in]{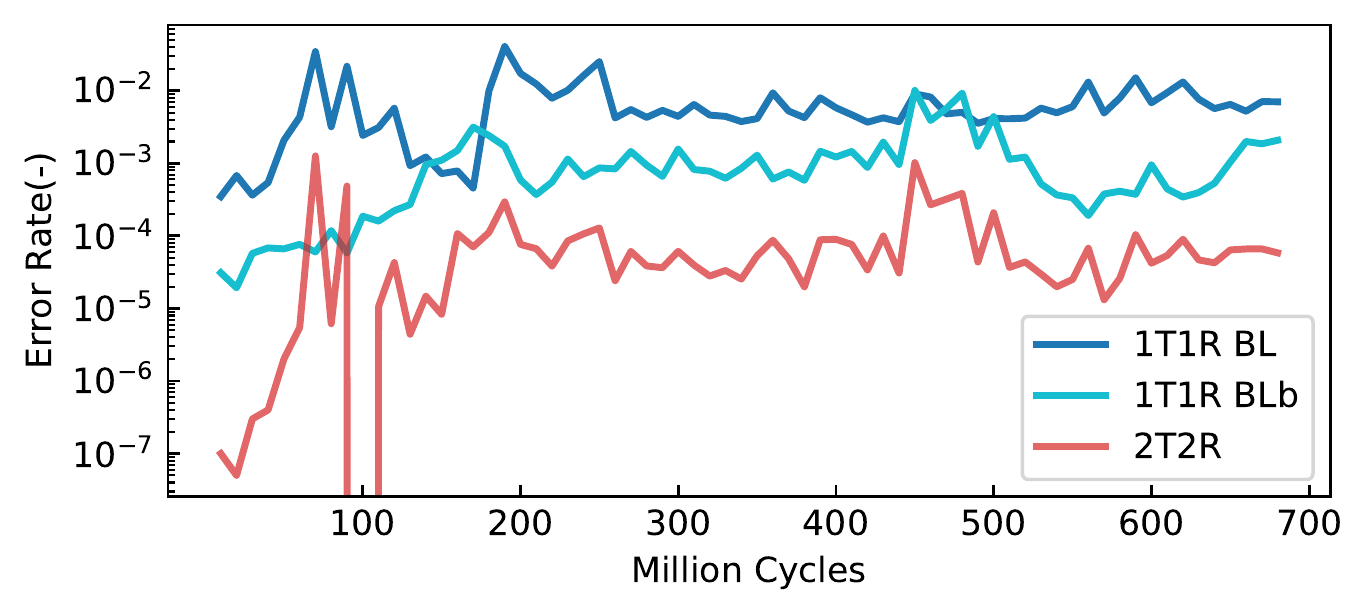}
    \caption{Mean bit error rate comparison between 1T1R \& 2T2R configuration over 10 million cycles as a function of the number of cycles that a device has been programmed.}
    \label{fig:expresult}
\end{figure}

Experimental results on the fabricated test confirm that the 2T2R approach reduces the amount of bit errors. Fig.~\ref{fig:expresult}, for example, shows  bit error rate measurements on a pair of devices within a kilobit memory array. 
The pair is reprogrammed 700 million times, alternating the programming  in high resistance/low resistance states and low resistance/high resistance states. The weight is measured after each programming event using the on-chip precharge sense amplifier in the 2T2R case, and is compared with direct sensing when single devices are used (1T1R). We see that the 2T2R error rate is two orders of magnitude below the 1T1R error.
More extensive experimental results on this test chip, involving various programming conditions and whole memory array measurements, are shown in \cite{bocquet2018,hirtzlin2019digital}. 
In particular, the results reported in these references indicate that the benefits of the 2T2R approach in terms of bit error rate reduction are similar to the one of formal single error correction of equivalent redundancy.

\begin{figure}
    \centering
    \includegraphics[width=3.0in]{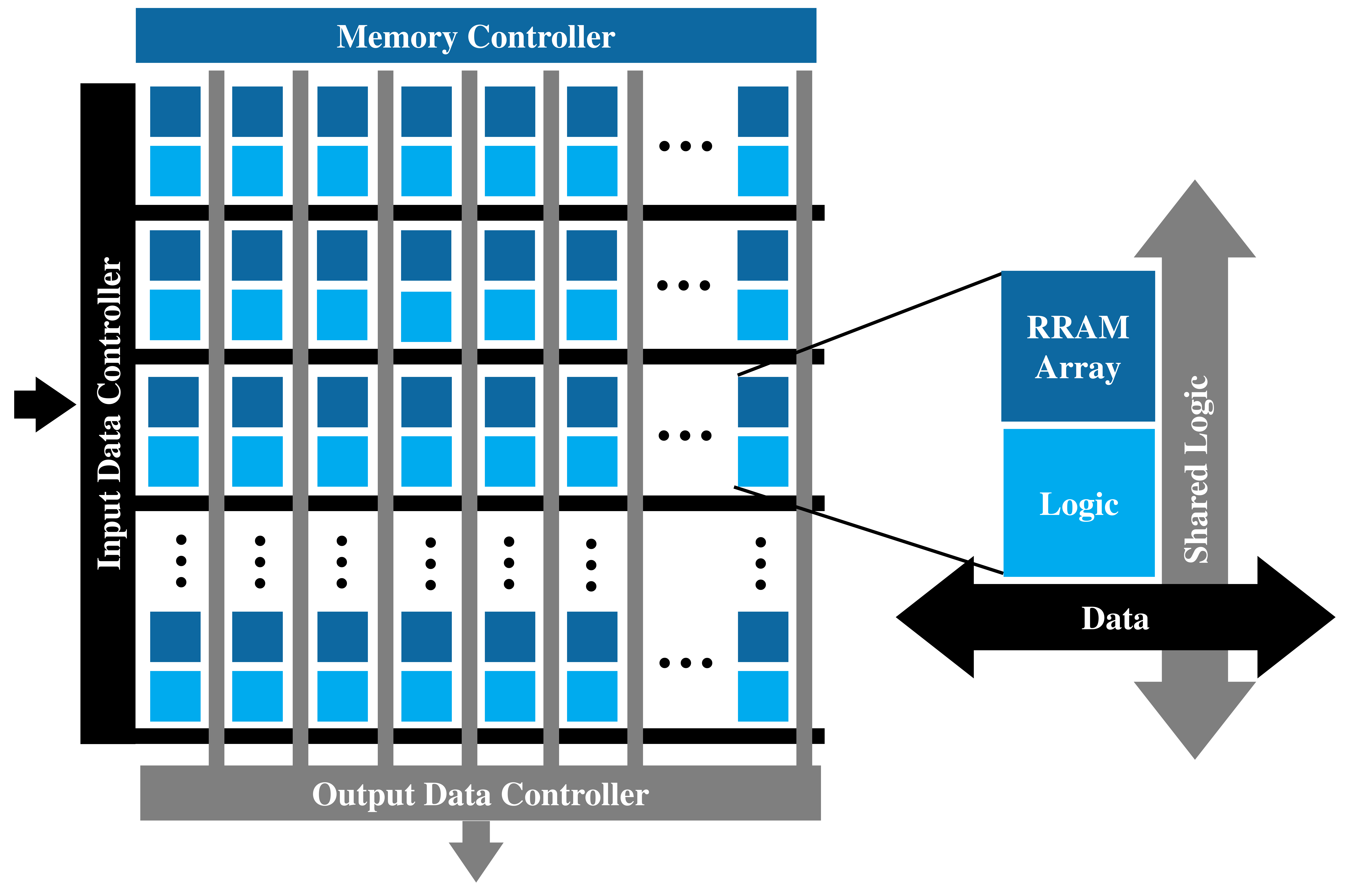}
    \caption{Schematic of the basic architecture for implementing fully connected BNN layer from in-memory computing basic blocks.}
    \label{fig:Architecture with RRAM}
\end{figure}

The memory arrays (Fig.~\ref{fig:die}) that we characterized can be used as basic building blocks for complete architectures implementing binarized neural networks. Fig.~\ref{fig:Architecture with RRAM} presents an architecture to implement fully connected layers, minimizing data movement, as described in detail in \cite{Hirtzlin2019b}.
The architecture incorporates RRAMs arrays with  sense amplifiers augmented with XNOR to perform binary multiplication; additional logic elements are added to perform popcount operations. The devices are programmed to neural network weights obtained by off-chip training. This programming occurs before the use of the inference circuit and is managed by a memory controller.
This type of architecture can be adapted for convolutional layers, with a key decision between minimizing data movement and data reuse: several works have investigated the implementation of convolutional layers from basic in-memory computing processing elements using static RAM \cite{valavi2018mixed} of emerging memories \cite{shafiee2016isaac,chi2016prime,giacomin2019robust}.

\section{Biomedical Time-Signals}

BNNs have  been mostly investigated for computer vision tasks. 
Their potential for low power hardware also makes them particularly attractive for medical signal analysis. 
For this reason, here, we investigate BNNs for that purpose,
and discuss resulting hardware implications.

\subsection{Electroencephalography (EEG) Task}

\begin{figure}
    \centering
    \includegraphics[width=2.5in]{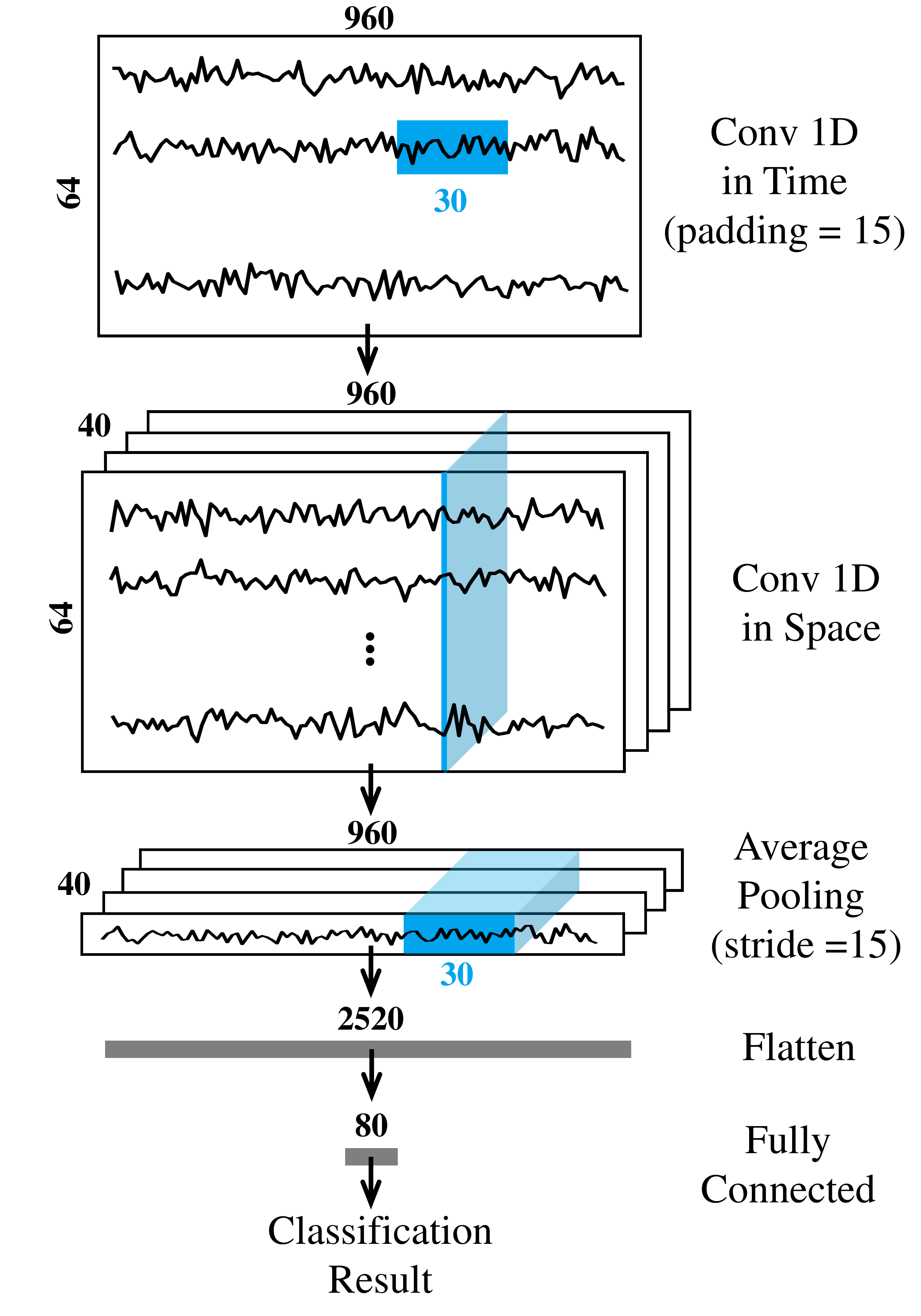}
    \caption{End-to-end EEG classification model proposed in \cite{Dose2018}. } 
    \label{fig:Dose_EEG_architecture}
\end{figure}

Electroencephalography is used to measure electric potentials from a human scalp surface, and has numerous applications, such as epilepsy diagnosis and prediction,
brain-computer interfaces or sleep stages monitoring.
As EEG suffers from  low signal to noise ratio,   complex analysis is usually required.
For this study we utilized data from the public EEG Motor Movement/Imagery Dataset \cite{BCI2000, PhysioNet}. Here we  consider a task of motor imagery:
a subject has to imagine moving their left or right fist, and
our neural network aims at detecting which of the two movements has been imagined,
based on six seconds time EEG measurements.
The complete set contains data of 109 subjects with $64$ electrodes sampled at 160~Hz.
We used a subset of $105$ subjects with $42$ trials of left-right
imaginary movements recorded. The four remaining subjects data were discarded as incomplete.
The only preprocessing step performed was per-channel normalization by subtracting
the mean and dividing by variance. 
As the dataset is not large, 
we used a relatively shallow neural network model, and we added small amplitude noise to each training sample for data-augmentation.
As a baseline, we exploit the end-to-end EEG classification neural network
model proposed in \cite{Schirrmeister2017, Dose2018} (Fig.~\ref{fig:Dose_EEG_architecture} and Table~\ref{table:Dose_EEG_summary}).
This  model consists of two convolutional layers followed by average pooling and a two-layer classifier.
The first layer processes individual signals as 1D images,
therefore performing 64 individual convolutions in time
(Fig.~\ref{fig:Temporal convolution}) over incoming EEG time-signals.
The second convolutional layer correlates obtained signals in space
simultaneously over all 64 channels.
We apply ReLU activations throughout the EEG model and replace them by $\textrm{sign}$ in a binarized setting.
The final $\textrm{softmax}$ layer is necessary only for training.
The real-weight version of this neural
network was reported to achieve $88\%$ accuracy \cite{Dose2018}.
To evaluate our binarization strategies, similarly to the baseline, we apply five-fold
cross-validation meaning that the dataset is partitioned into
five non-overlapping validation subsets not seen during the training.
We report an average over five experiments
where we train a new model from scratch over 1000 epochs with the Adam method \cite{kingma2014adam} in each experiment.

\begin{table}[h]
  \begin{center}
    \caption{EEG classification network architecture from \cite{Dose2018}.
    }
    \begin{tabular}{l c r r}
      \hline
          & Kernels & Padding & Output shape \\
      \hline
       Conv 40 & $30 \times 1$ & 15 & $961 \times 64 \times 40$ \\
       Conv 40 & $1 \times 64 \times 40$ & No & $961 \times 1 \times 40$ \\
       Avg. pool & $30 \times 1$ & No & $63 \times 1 \times 40$ \\
       Flatten & - & - & 2520 \\
       FC 80 & - & - & 80 \\
       Softmax & - & - & 2 \\
      \hline
    \end{tabular}
    \label{table:Dose_EEG_summary}
  \end{center}
\end{table}

\subsection{Electrocardiogram (ECG) Task}

Electrocardiogram signals measure electrical changes
as a result of cardiac muscle depolarization,
and are used e.g. for diagnosing different kinds of arrhythmia. 
Typically, ECGs are recorded using 12 electrodes, which need to be properly positioned:
inverting any pair of electrodes may lead to wrong diagnosis.
We here consider the task of detecting such electrode inversion.
We used a dataset from the Challenge Data competition \cite{electrode_inversion_detection}. The dataset contains 1000 trials of three second recordings performed at 250~Hz.
Our custom CNN model is summarized in Table~\ref{table:ecg_network}.
Each convolution/linear layer is followed by batch normalization and nonlinear activation. We replace $\textrm{hardtanh}$ activation by a
$\textrm{sign}$ in a binarized setting.
In addition, we also perform batch normalization of the input data.
We train the model using the Adam optimizer \cite{kingma2014adam} over 1000 training epochs and, as in the EEG task, we perform five-fold cross-validation five times.
To address overfitting, we employ a dropout regularization with keep probability $0.95$ within convolution layers and $0.85$ within the classifier.

\begin{table}[h]
  \begin{center}
    \caption{ ECG classification network architecture. 
    }
    \begin{tabular}{l l l l}
      \hline
          & Kernels & Padding &  Output shape \\
      \hline
       Conv 32 & $13 \times 1 \times 12$ & No & $738 \times 1 \times 32$ \\
       Max. pool & $2 \times 1$ & No &  $369 \times 1 \times 32$ \\
       
       Conv 32 & $11 \times 1 \times 32$ & No &  $359 \times 1 \times 32$ \\
       Max. pool & $2 \times 1$ & No &  $179 \times 1 \times 32$ \\
       
       Conv 32 & $9 \times 1 \times 32$ & No  &  $171 \times 1 \times 32$  \\
       
       Conv 32 & $7 \times 1 \times 32$ & No &  $165 \times 1 \times 32$ \\
       
       Conv 32 & $5 \times 1 \times 32$ & No &  $161 \times 1 \times 32$  \\
       
       Flatten & - & - &  5152 \\
       FC 75 & - & - &  75 \\
       Softmax & - & - &  2 \\

      \hline
    \end{tabular}
    \label{table:ecg_network}
  \end{center}
\end{table}

\begin{table}[h]
  \begin{center}
    \caption{Accuracy comparison of  CNN with real weights, binarized CNN (BNN), and CNN where only the fully-connected part was binarized. In parentheses, number of filter augmentations.
    }
    \label{tab:NNaccuracies}
    \begin{tabular}{l l l l}
      \hline
      Task & Real-weight NN & BNN & Bin. Classifier \\
      \hline
      EEG & 88\% \cite{Dose2018} & \makecell[l]{{84.6}\% ($1 \times$) \\ {86}\% ($11 \times $)} & {87}\% ($1 \times$)  \\
            ECG & 96.3\% &  \makecell[l]{92.1\% ($1 \times$) \\  94.9\% ($7 \times $)} &  95.9\% ($1 \times$)\\
      ImageNet Top-1 & 70.6\%  \cite{Howard2017} & 54.4\%~($4 \times $)~\cite{Phan2019} & 70\% ($1 \times$)\\
      ImageNet Top-5 & 89.5\%  \cite{Howard2017} & 77.5\%~($4 \times $)~\cite{Phan2019} & 89.1\% ($1 \times$)\\
      \hline
    \end{tabular}
  \end{center}
\end{table}

\begin{table}[h]
  \begin{center}
      \caption{Model memory usage comparison between different tasks and savings with classifier binarization.}
    \begin{tabular}{l l l l l}
      \hline
          \thead{Model} & \thead{Total \\ params} & \thead{Classifier \\ params} & \thead{Model size \\ 32-bit / 8-bit} & \thead{Bin classif. \\ saving \%} \\
      \hline
       EEG & 0.31M & 0.2M & 1.17MB / 305KB & 64\% / 57.8\% \\
       ECG & 0.31M & 0.27M & 1.17MB / 305KB & 84\% / 75.8\% \\
       ImageNet & 4.2M & 1M & 16.2MB / 4.1MB & 20\% / 7.3\% \\
             \hline
    \end{tabular}
    \label{table:memory_usage_comparison}
  \end{center}
\end{table}

\subsection{Results}

\begin{figure}
    \centering
    \includegraphics[width=3.0in]{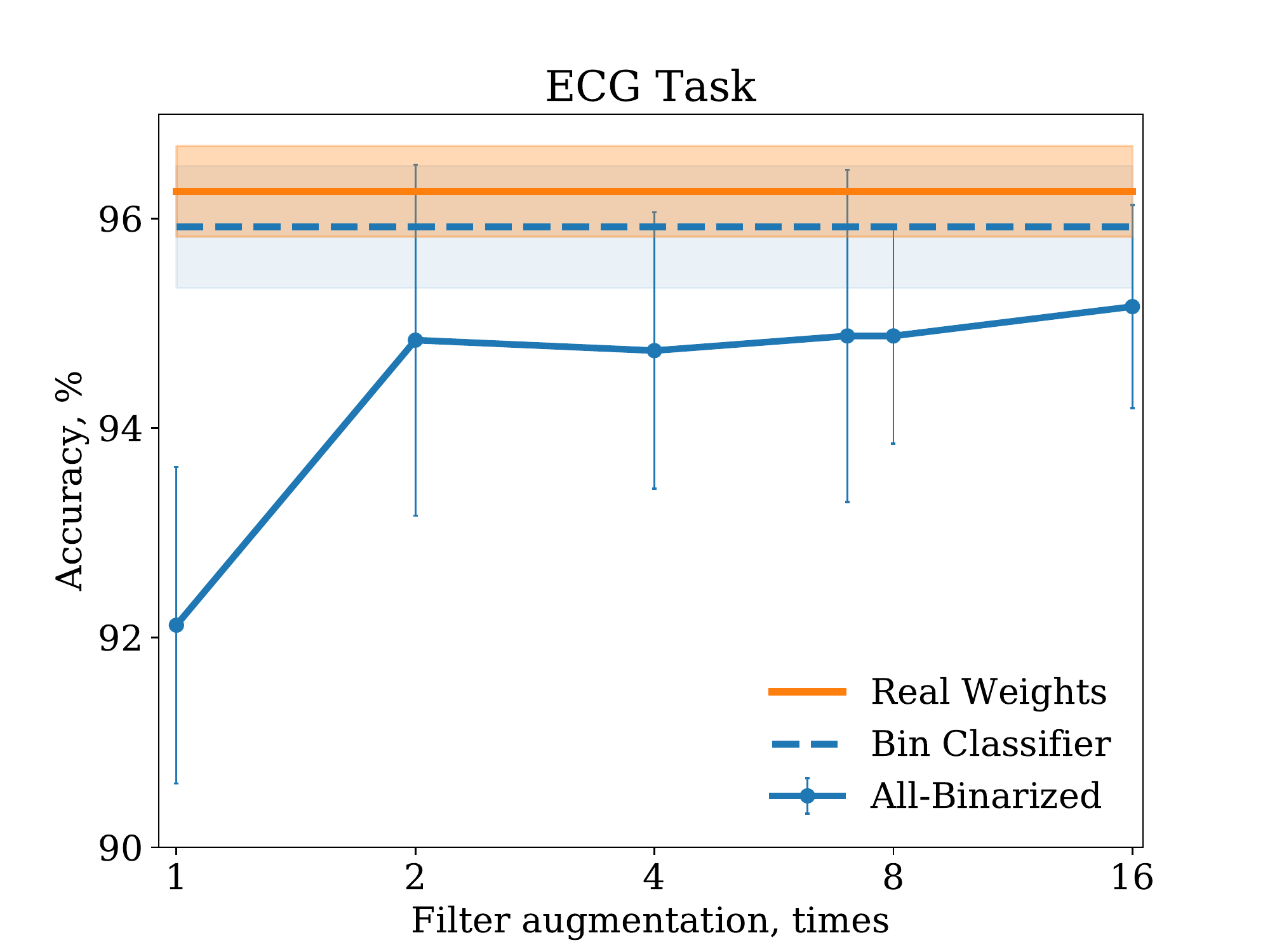}
    \caption{Cross-validated accuracy on the ECG task, for different models. BNN accuracy is improved when increasing the number of convolution filters. 
    }
    \label{fig:ecg-filter-aug}
\end{figure}

We performed simulations on the ECG and EEG
neural network architectures in three cases: 
real weights (32-bit floating point), fully binarized neural network, and a mixed situation (binarized classifier) where fully-connected layers are binarized, while convolutional layers remain real.
Fig.~\ref{fig:ecg-filter-aug} shows a detailed result in the ECG case.
We see that a fully binarized neural network on average performed worse than a real weight neural network if an equivalent number of filters is used ($92.1\%$ vs. $96.3\%$ accuracy).
The introduction
of more filters allows enhancing  the accuracy of the BNN; however, the accuracy of the real network is not reached.
Fig.~\ref{fig:ecg-filter-aug} also shows that the network where only the classifier has been binarized is able to match the accuracy of the real neural network (within error bar), without any augmentation of the number of filters.
Similar results were obtained with the EEG task, as reported in  Table~\ref{tab:NNaccuracies}.

These results have interest for hardware development.
For both ECG and EEG models, most of the weights reside in fully-connected classifier layers, and the strategy of binarizing only the classifier has therefore high memory benefits. 
A detailed analysis of the memory savings of the different strategies is presented in Table~\ref{table:memory_usage_comparison}.
For example, the EEG model requires 0.31M total parameters, occupying 1.17MB of memory;
0.11M of parameters (406 KB) are in convolutional layers and
0.2M (789 KB or $66\%$ of total parameters), in the classifier.
Therefore, binarizing only the classifier can save
about $64\%$ compared to the 32-bit model.
If we compare with a neural network quantized to eight-bit numbers, the memory saving would be $57.8\%$ of memory. 
Even better memory savings are obtained with the ECG model: by binarizing only the classifier,
the memory reduction is $84\%$ compared to the
original 32-bit model.
If we used an eight-bits numbers quantized network as a reference, we would save $75.8\%$ memory.
The strategy of only binarizing the classifier also achieves better accuracy than a fully binarized network using an equivalent amount of memory: the binarized classifier model accuracy is
by $1\%$ better for EEG and by $1\%$ for ECG,
compared to those with all-binarized network of equivalent
number of bits (11 times BNN convolution filter augmentation for EEG, 7 times augmentation for ECG BNN model).
If we assume that convolutional layers can be quantized to eight-bits precision, the accuracy gap is then $2.3\%$ for EEG ($3 \times$ augmentation) and $1\%$ for ECG ($2 \times$).

\section{Partial Binarization on  MobileNet}

\begin{figure}
    \centering
    \includegraphics[width=3.0in]{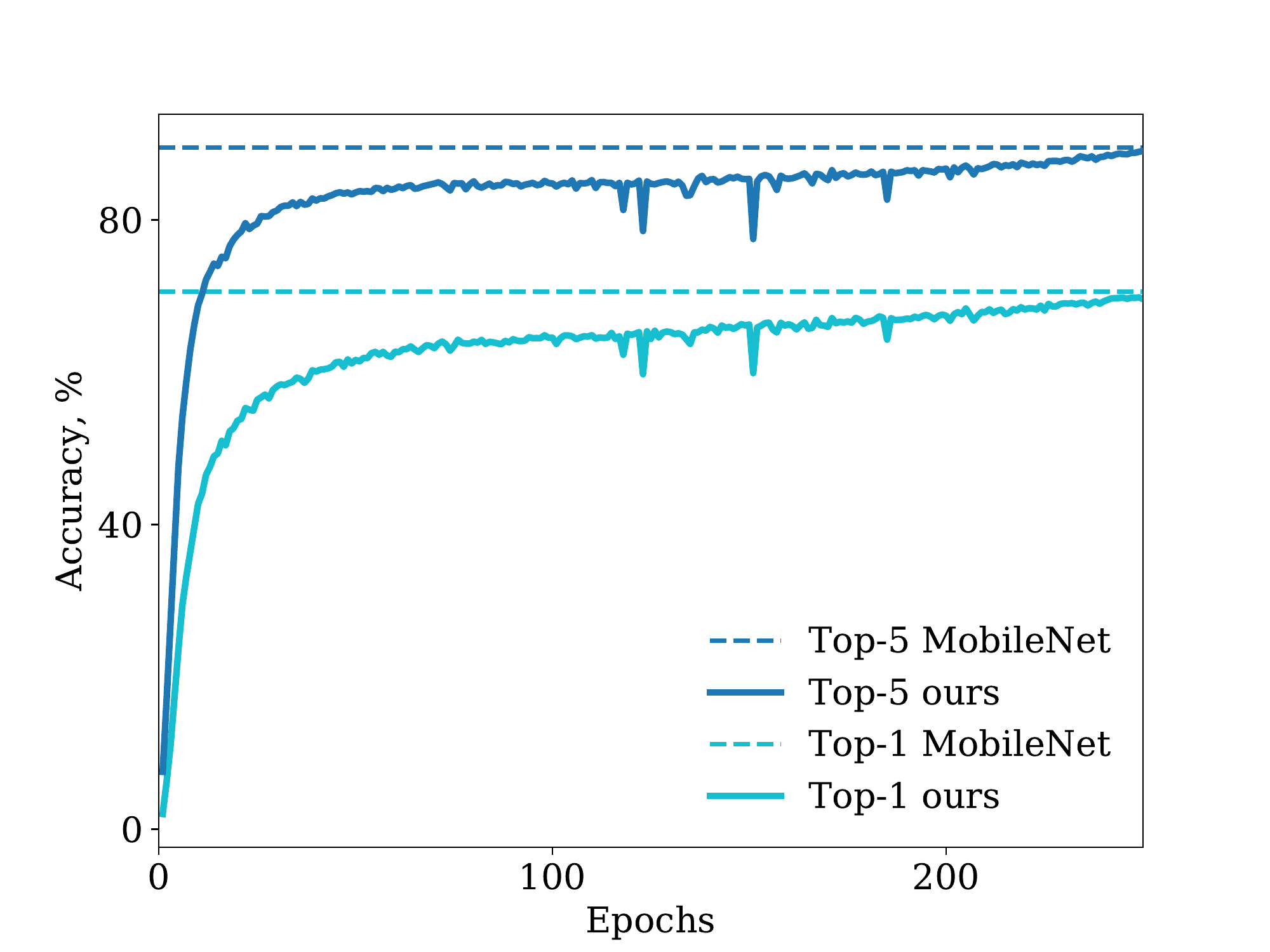}
    \caption{Training modified MobileNet-224 model with a binarized classifier on the ImageNet 1K challenge.}
    \label{fig:mobilenetv1}
\end{figure}

Machine learning can be effectively applied to analyze
data obtained by medical imaging, and  we envision that binarized neural networks
could  be successful in such contexts.
To evaluate the applicability of our approach to vision tasks on neural networks optimized for mobile applications,  we evaluate our approach of classifier
binarization on the generic ImageNet dataset
 \cite{ILSVRC15},
containing 1.2 million images representing 1000 classes.
We utilize MobileNet~V1, a compact network specifically designed
for mobile devices with less computing power \cite{Howard2017}.
MobileNet architecture is characterized by
replacing most of standard convolutions with
depthwise-separable convolutions,
which require less computations.
In this work, we replaced the fully-connected classifier
with a binarized classifier of two layers.
We trained the network from scratch with stochastic gradient
descent method in $255$ epochs.
We achieved Imagenet Top-1 accuracy close to the original MobileNet-224 (70.6\% vs. 70\% bin classifier)
and equally, Top-5 accuracy (89.5\% vs. 89.1\% bin classifier) (Fig.~\ref{fig:mobilenetv1}), whereas fully binarizing Mobilenet~V1 is associated with accuracy degradation (Table~\ref{tab:NNaccuracies}).
This result further confirms that classifiers binarize more naturally than convolutional layers. 

The MobileNet-224 model has 4.2M parameters (16.2MB),
3.2M parameters (12.28MB) of which are used by
convolutional layers and 1M (4.05MB), by the original
single-layer classifier (Table~\ref{table:memory_usage_comparison}).
Therefore, the classifier occupies about 24\% of
memory. As a binarized classifier
we use two layers of 5.7M binary parameters (696KB),
therefore this allows us to save
about 20\% of memory
compared to the network with all 32-bit weights.
In case if we used 8-bit numbers as a reference, we would
still spare about 7.3\% memory.

\section{Conclusion}

In this work, we highlighted that binarized neural network can be a road to implement particularly efficient neural networks hardware.
We introduced an implementation, validated by measurements on a fabricated test chip using hafnium oxide resistive memory. Our implementation is designed around the principles of in-memory computing, limiting the amount of data movement, and avoiding error correcting code altogether.
As BNNs have mostly been evaluated on vision tasks, and medical signal analysis is believed to be an essential application for highly efficient AI chips, we evaluated BNNs on two sample ECG and EEG signal analysis task.
We report that all-binarized neural networks can be a tool
to reduce memory requirements. On the other hand, if keeping
the highest accuracy is vital, we propose an alternative route where only the classifier part of the neural network is binarized. Moreover, as these neural networks are dominated by classifier, non volatile memory requirement can be considerably reduced by classifier binarization.
We also evaluated the strategy of binarizing only the classifier on a vision task, on an architecture optimized for modest memory requirement (MobileNet~V1). The resulting neural network matches the original MobileNet~V1 network accuracy on ImageNet large-scale benchmark, however the nonvolatile memory benefits are smaller than in the EEG and ECG tasks, as MobileNet~V1 is dominated by convolutions.

These first results are very encouraging to reduce efficiently memory requirement of edge devices and thus to obtain low energy hardware. This is particularly useful for medical applications where little energy is available. The results also highlight
that such hardware development should be pursued in co-development with neural network architecture and in conjunction with application development, as the most efficient approaches can be considerably task-dependent. 

\bibliography{IEEEabrv,db}
\bibliographystyle{IEEEtran}

\end{document}